# Current induced magnetization dynamics in current perpendicular to the plane spin valves

M. Covington[a], M. AlHajDarwish[b], Y. Ding[c], N. J. Gokemeijer, and M. A. Seigler

*Seagate Research, 1251 Waterfront Place, Pittsburgh, PA 15222*

We observe magnetization dynamics induced by spin momentum transfer in the noise spectra of current perpendicular to the plane giant magnetoresistance spin valves. The dynamics are observable only for those combinations of current direction and magnetic configuration in which spin transfer acts to reorient the free layer magnetization away from the direction set by the net magnetic field. Detailed measurements as a function of magnetic configuration reveal an evolution of the noise spectra, going from a spectrum with a well-defined noise peak when the free layer is roughly collinear with the pinned layer to a spectrum dominated by $1/f$ noise when the free layer is in an orthogonal configuration. Finally, the amplitude of the corresponding resistance noise increases rapidly with increasing current until it saturates at a value that is a substantial fraction of the magnetoresistance between parallel and antiparallel states.

## I. Introduction

Magnetic fields generated by moving charges can exert a torque on a magnetic moment. Recent research, however, has confirmed the existence of an entirely new effect called spin momentum transfer that can also manipulate the magnetization of a ferromagnet via an electrical current. The heart of the effect lies in the exchange of spin angular momentum between conduction electrons and the magnetic moment of a ferromagnet. This interaction can lead to a net torque acting on a ferromagnet and the possibility of an electrical current directly inducing magnetization dynamics. There are numerous theoretical models of this spin torque [1-6]. Despite the differing microscopic approaches, all of these models derive a spin torque that is qualitatively different than the classical torque exerted by a magnetic field. One of the key distinguishing features of spin torque is that, depending on current direction, it can have a substantial component that either opposes or enhances the magnetic damping. Theory predicts that this property can lead to a state of persistent magnetization precession driven by a dc current [1, 2, 7-9].

Experimental observations of spin transfer have come from measurements of either nanopillar or point contact structures. By using the magnetoresistance to infer the magnetic state of the device, experiments have demonstrated that spin transfer can reproducibly reorient the magnetization of a small ferromagnetic element between two bistable configurations [10-14]. Further dc measurements have inferred states of current induced magnetization precession that are consistent with spin transfer [10, 15-17]. More recent experiments have reported signatures of these spin transfer induced precessional states in the frequency domain [18, 19].

Spin momentum transfer is of interest for both fundamental and practical reasons. Studying this effect can further expand the understanding of microscopic spin dependent interactions between electrons. Spin dependent scattering is the mechanism behind many magnetoresistance phenomena [20]. In addition, recent work that is complementary to the work on spin transfer has shown that conduction electrons can have a significant impact on the damping of magnetization dynamics in ferromagnetic thin films [21]. From a practical standpoint, an understanding of spin transfer is important for the application of current perpendicular to the plane (CPP) giant magnetoresistance (GMR) devices. Hard disk drives currently use GMR read sensors in which current flows in the plane of the thin film multilayer (CIP), but CPP GMR sensors are considered to be a serious alternative for future drives. CPP GMR will encounter the same Johnson noise and thermally activated magnetic noise, or mag-noise [22-24], as CIP GMR devices. However, spin transfer is an additional noise mechanism in CPP GMR devices that must be understood.

In this article, we present experimental observations of spin transfer induced magnetization dynamics in the frequency domain. Our work builds upon the work from other groups that have observed similar results [18, 19]. In addition to extensive measurements of the current dependence of the dynamics, we exploited the flexibility of magnetic biasing allowed by CPP spin valves and have made detailed measurements as a function of magnetic configuration. In section II, we describe the CPP spin valves and the measurement apparatus used to study the devices. Section III presents the experimental results of the device characterization and current induced noise measurements. Finally, we summarize our data in section IV.

## II. Experiment

We study current-induced magnetization dynamics by measuring the noise spectra of CPP GMR spin valves. The devices have a layer structure from bottom to top of IrMn90/CoFeX/Ru10/CoFeX/Cu22/CoFe30, where the numbers indicate the layer thickness in Å. Devices from two wafers were tested in which X=30 Å or 40 Å. The data in this paper are from a device where X=40 Å and are representative of the behavior observed from both wafers. The thin films were sputter deposited, and the devices were fabricated using standard wafer processing. The CPP pillars have a rectangular cross section with dimensions of approximately 100 nm x 200 nm and are connected by four electrical leads.

The synthetic antiferromagnet (SAF) structure CoFe/Ru/CoFe strongly couples the two CoFe layers antiferromagnetically via Ruderman-Kittel-Kasuya-Yosida (RKKY) coupling [25], and the lower and upper CoFe layers within this structure are referred to as the pinned layer (PL) and reference layer (RL), respectively. Saturation fields of ~5 kOe have been measured for sheet films of similar unpinned SAF structures. The combined layers of IrMn/CoFe/Ru/CoFe are a stable magnetic structure that serves to fix the magnetizations of the PL and RL within the SAF. The GMR from the spin valve is then a measure of the relative angle between the fixed magnetization of the RL and the freely rotatable magnetization of the upper CoFe layer, referred to as the free layer (FL) [20]. All subsequent references to the PL, RL, and FL in this article will refer to the magnetizations of the corresponding layers, even without an explicit citation as such in the text.

All measurements were performed with an apparatus that can make electrical connection to the on-wafer devices with either tungsten pins or microwave probes, apply magnetic fields via a quadrupole electromagnet, and heat the wafer above room temperature. DC electrical measurements were acquired using standard two- and four-probe geometries. Frequency domain noise measurements used 50 $\Omega$ microwave circuitry between the device and spectrum analyzer [26]. A current source and voltmeter, isolated from the microwave components through a bias tee, were used to apply a dc current and simultaneously measure the dc voltage. A 40 dB amplifier was also used that had a 0.1 to 8 GHz bandwidth and a voltage noise equivalent to a 70 $\Omega$ resistor at 293 K referenced at the input. The spectrum analyzer was calibrated prior to measurements with two sources that output white noise 6 dB and 15 dB larger than that of a 50 $\Omega$ resistor at 293 K. A calibration curve was generated from the known and measured noise and was used to correct the noise spectra. This calibration procedure only corrects for signal loss up to the end of the coaxial cable connected to the microwave probes. However, separate measurements of a wide bandwidth short, open, and 50 $\Omega$ load on a special calibration chip indicate that the additional loss coming from the probes and on-wafer device is less than 0.5 dB and essentially independent of frequency.

### III. Results

*Device characterization*

The magnetic and magnetoresistive properties of the CPP spin valves were characterized by measuring the four-probe resistance versus field. Figure 1 shows typical resistance versus field data from a CPP spin valve for fields applied along the easy axis of the device. The magnetoresistance between the FL and RL is much larger than that originating between the RL and PL [27] and the resistance will hereafter be used to infer the relative magnetic configuration between only the FL and RL.  The PL and RL will also be assumed to be antiparallel for all of the experimental conditions presented.  The sharp jumps in resistance around ±300 Oe are due to the reversal of the FL magnetization as it aligns along the direction of the magnetic field.  The resistance change around +1 kOe comes from the reversal of the entire pinned SAF structure, where the RL and PL remain antiparallel but the RL switches between parallel and antiparallel alignment with respect to the FL.

There are three factors that lead to magnetic anisotropy in the device, and the resulting uniaxial and unidirectional anisotropies are all oriented along the long axis of the CPP pillar.  First, the CoFe layers have a uniaxial anisotropy of $H_K$ ~10 Oe induced by an aligning magnetic field during thin film deposition.  Second, the antiferromagnetic IrMn layer adds a ~500 Oe unidirectional anisotropy field to the PL.  The third, and most dominant, factor comes from the shape anisotropy of the rectangular pillar, with the easy axis lying along the long axis.  The device shape induces an effective uniaxial anisotropy in the FL and helps to stabilize the SAF above and beyond that achieved at the sheet film level.  Estimates assuming a uniform magnetization yield uniaxial anisotropy fields of approximately 400 and 500 Oe for individual CoFe layers that have a 100 nm x 200 nm cross section and are respectively 30 and 40 nm thick.  Applying this same assumption of uniform magnetization to the experimental spin valve geometry, this model further predicts that the magnetic fields that saturate the PL, RL, and FL magnetizations into a parallel state differ by approximately 1.5 kOe for fields applied along the long and short axes of the structure.  Hence, magnetostatic fields favor antiparallel alignment between adjacent magnetic layers and, within this antiparallel configuration, further reinforce the magnetizations to lie along the long axis of the pillar.  We also argue that these magnetostatic coupling fields are contributing factors to the small decrease in resistance between 1 and 2.5 kOe, which is interpreted to be due to increasing uniformity of the FL magnetization as it achieves a more fully parallel state with respect to the RL.

Figure 1b focuses on the characteristics of the FL minor loop as represented via the resistance, $\Delta R$, with respect to the low resistance state at an easy axis field of -600 Oe.  For small bias currents and no hard axis magnetic field, the GMR ratio is 1.8% and $\Delta R \cdot A$ =1.53 mΩ-μm$^2$ for an assumed cross sectional area, $A$, of 0.02 μm$^2$. The $\Delta R$ decreases for larger bias currents or when a dc hard axis field is simultaneously applied. This decrease in $\Delta R$ for large currents is likely due to curling of the FL and RL magnetizations induced by the self-fields generated from the current flowing through the CPP pillar, which consequently limits the ability to achieve fully parallel or antiparallel magnetic states between the FL and RL. The decrease in $\Delta R$ resulting from the application of a hard axis field can be accounted for by simple canting of the FL magnetization away from the easy axis. Lastly, the offset of the FL reversal from $H_{EA}$=0 observed with 18 mA (curve 3) is due to self-fields generated by current flowing in the leads connected to the CPP device. The FL reversal occurs at either positive or negative easy axis fields depending on which of the two top leads are used to pass current through the CPP pillar. The direction of the shift is consistent with the direction of the in-plane field produced by current flowing from the CPP pillar into the top leads, and the offset from $H_{EA}$=0 is consistent with order of magnitude estimates of the self-fields generated by the current. These self-fields also promote non-hysteretic FL reversal.

*Noise measurements*

The primary result in this article is the observation of noise from CPP spin valves that is a function of magnetic configuration and bias current direction and amplitude, which we argue arises from magnetization dynamics induced by spin momentum transfer. Figure 2 summarizes the characteristics of CPP spin valve noise spectra. Figure 2a shows examples of noise spectra acquired for three different relative magnetic configurations between the FL and RL as set by the easy axis field [28]. An easy axis field of 60 Oe biased the device into a high resistance state, -10 Oe biased it into a low resistance state, and 34 Oe biased it into an intermediate resistance state. The magnetoresistance implies that the FL magnetization is canted at a finite angle with respect to the RL magnetization at 60 and -10 Oe, but the corresponding high and low resistance states will be referred to as antiparallel and parallel, respectively, for the sake of brevity. The resistance at 34 Oe implies that the FL and RL magnetizations are nearly orthogonal.

In Figure 2a, the device was biased with a +20 mA current. Positive current is defined as flowing from the bottom to the top of the CPP pillar, so electrons are flowing from the FL to the RL. To enable the FL to be biased at arbitrary angles with respect to the RL, a constant 1375 Oe hard axis field was also applied. This hard axis field value was determined by the device magnetics so that the device exhibited non-hysteretic rotational FL reversal rather than hysteretic switching during the noise measurements. For these conditions, the noise in the antiparallel state is below the noise floor of the electronics. However, for the same current, the noise exhibits a dramatic increase in amplitude and change in spectral content as the FL is rotated towards the parallel state. The spectrum is dominated by noise exhibiting a $1/f$ functional form when in the orthogonal state. But, the low frequency noise decreases and a well-defined peak develops when the device is in the parallel state [29]. For this latter case, there is still spectral power distributed over a broad range of frequencies, but the functional dependence below the peak frequency is no longer $1/f$ and is instead closer to white noise. Similar behavior is observed for the opposite current direction, but for that case the noise is largest when the magnetic configuration is either orthogonal or in the antiparallel state.

A compendium of noise spectra is shown in Figures 2b and 2c. The three dimensional plots are constructed from individual noise spectra measured at 20 Oe intervals of the easy axis field. Also shown is the two-probe dc resistance that reflects the magnetic configuration. These figures illustrate the strong correlation of the noise with current direction and magnetic configuration. For positive currents, where the spin transfer torque drives the FL towards the antiparallel state, the device noise appears when the FL magnetization is in or near the parallel state. For negative currents, where spin transfer favors the parallel state, the noise is observed when the FL magnetization is in or near the antiparallel state. Whenever the current direction is such that spin transfer reinforces the FL magnetization along the same direction as that set by the net magnetic field, the device noise falls below the electronics noise.

Additional noise measurements have been made as a function of current and hard axis magnetic field, and from devices with different cross-sectional areas. Some of the key trends are as follows. The noise spectra exhibit the same qualitative features as the hard axis field increases from 1000 to 1625 Oe, albeit with a decrease in noise amplitude and increase in the peak frequencies. The noise spectra exhibit the same characteristics regardless which of the two possible directions the hard axis field is applied. Examples of peak linewidths in Figure 2 indicate that $\left( \Delta f / f_{peak} \right)$ is approximately 0.017 for +20 mA and 0.08 for -20 mA, where $\Delta f$ is the full width at half maximum and $f_{peak}$ is the peak frequency. As the current amplitude increases, the amplitude of the noise also increases. The current densities used for these experiments are on the order of $1 \times 10^8$ A/cm$^2$. The devices are stable and reproducible when biased with these current densities, which are roughly an order of magnitude less than those used in experiments with point contacts [15]. The peak frequency typically decreases with increasing current, although separating shifts due self-field effects of the bias current from possible shifts due to spin transfer is difficult to distinguish in these devices. The noise amplitude for a given current is larger for higher resistance devices with smaller cross-sectional areas. Lastly, while different devices may exhibit detailed differences in characteristics, the same correlation between the noise, current, and magnetic configuration is consistently observed.

Since the noise is dependent on bias current, we hereafter will assume that the data are resistance noise [30] due to a combination of FL magnetization dynamics and GMR. We will also neglect fluctuations of the PL or RL magnetizations within the SAF. We argue that this latter assumption is valid because the combined effect of the interlayer coupling across the Ru layer and the exchange coupling to the IrMn will lead to a substantially stiffer magnetic system with much smaller fluctuation amplitude and much higher precessional frequencies that are likely beyond our detection capability. A further justification is that the data in Figure 2 are qualitatively consistent with a pinned RL and a fluctuating FL.

The existence of persistent microwave oscillations for the current directions and magnetic configurations shown in Figure 2 is a hallmark of spin transfer torque, and the corresponding spectral peak is consistent with theoretical predictions [1, 2, 7-9]. But, the existence of low frequency noise is somewhat unexpected. The correlation of the low frequency noise with current direction and magnetic configuration indicates that it too is being driven by spin transfer. Low frequency noise similar to the results presented in this paper also appears to be a common occurrence. The distribution of spectral noise power over a broad frequency range and specific instances of noise exhibiting either $1/f$ or Lorentzian behavior have been observed in comparable CPP structures [18, 19, 31]. Furthermore, recent numerical micromagnetic modeling studies of the Landau-Lifshitz-Gilbert (LLG) equation of motion with spin torque included indicate that such low frequency noise can indeed result from spin transfer. One study has shown that the combination of a random thermal force and spin torque can drive the magnetization to randomly jump between degenerate precessional orbits [8]. Although the spectral content of the dynamics was not shown, this hopping between orbits did introduce lower frequency components to the in-plane magnetization components. Similar numerical work has also predicted the existence of low frequency noise [9]. Nevertheless, a more complete theoretical study is needed to distinguish between the various effects of spin transfer, random thermal forces, and magnetic inhomogeneity and the influence each effect has on the low frequency noise.

More detailed measurements of the noise in the vicinity of the FL reversal, in which spectra were measured at 2 Oe increments of the easy axis field, reveal changes in noise amplitude as a function of magnetic configuration. The spectra exhibit the same qualitative features as shown in Figure 2, but the data are presented in a more compact form in terms of the RMS voltage noise. Figure 3 shows a typical example of the dc resistance and RMS voltage noise as a function of magnetic configuration. In addition to the asymmetry in noise amplitude about the FL reversal, the noise in Figure 3b also exhibits a peak in the vicinity of the FL reversal, where the FL is close to an orthogonal configuration with respect to the RL. The data in Figure 3b were calculated by integrating the measured spectra over the full experimental bandwidth of 0.1 to 8 GHz. Since the spectral content is dependent on the FL configuration, Figure 3c shows the RMS voltage noise integrated over a low frequency bandwidth of 0.1 to 4 GHz and a high frequency bandwidth of 4 to 8 GHz. The high frequency noise captures the evolution of the spectral peak and closely tracks the dc resistance, which is consistent with expectations of a spin transfer induced state of magnetization precession. In contrast, the amplitude of the low frequency noise is largest near the FL reversal.

Although we have yet to do micromagnetic simulations of our devices, we attribute the variation of the low frequency noise with easy axis field to a combination of at least three well-established effects. First, the asymmetry in noise amplitude between the low and high resistance states indicates that spin transfer is driving the noise, which accounts for why the device is quiet when the FL magnetization is in the parallel state and noisier in the antiparallel state. Second, the dependence of the GMR sensitivity, $\partial R/\partial \theta$, on the relative angle, $\theta$, between the FL and RL magnetizations can partly describe the peak in noise amplitude near the intermediate resistance state. For a constant RMS fluctuation of $\theta$, the RMS voltage noise will peak near the orthogonal magnetic state, where the GMR sensitivity is largest. Third, an increase in the net magnetic field acting on the FL can stiffen the FL magnetization, thereby suppressing the angular deviations of the FL about its average value. The increase in peak frequency as the easy axis field drives the FL magnetization away from the reversal suggests that the net field is increasing [18, 19], which will suppress the noise.

Examples of the current and field dependence of the RMS resistance noise are shown in Figure 4. These data are defined by the equation $R_{RMS} = \left[\left(V_{RMS}^2 - V_0^2\right)/I_b^2\right]^{1/2}$ and correspond to noise integrated over 0.1 to 8 GHz. The electronics noise, $V_0^2 = (88 \ \mu V)^2$, is subtracted from the total measured noise, $V_{RMS}^2$, in order to get the RMS voltage noise of the device itself, which is then converted from voltage noise to resistance noise via the bias current, $I_b$. Note that shifts in the FL reversal fields due to self fields from the bias current consequently lead to current dependent shifts in the transition between low and high noise states in the noise versus field curves. Interestingly, for positive current flow, the RMS resistance amplitude saturates at approximately 14 mΩ. This saturation value is a large fraction of the $\Delta R = 62$ mΩ measured for +20 mA and a 0 Oe hard axis field [32], but it is still consistent with the hypothesis that the noise originates from FL dynamics and GMR. This holds true even when integrating the $1/f$ noise over a larger bandwidth and will be discussed further below.

Coincident with the saturation in the resistance noise is the broadening, in Oe, of the FL reversal. Figures 1 and 2 show typical characteristics of the FL minor loop acquired with a 1375 Oe hard axis field and positive bias currents of +20 mA and below.

The resistance changes gradually with easy axis field except for a rapid change in resistance over a very small field range. This abrupt reversal in the FL magnetization occurs over a field range of approximately 12 Oe or less for currents that are less than or equal to +20 mA. However, this transition broadens when the current increases above +20 mA and, as an example, is approximately 70 Oe at +23 mA. This is similar to previous work [10, 15-19] and reflects how the regime of strongly driven precession can modify the average magnetic configuration of the device.

The current dependence is also plotted in an alternate form in Figure 5. These data have been integrated over the full experimental bandwidth of 0.1 to 8 GHz but are also representative of the behavior exhibited by the integrated noise from the 0.1 to 4 GHz and 4 to 8 GHz bandwidths. The shifts in the FL reversal field are accounted for by plotting the noise measured at a fixed field value away from the midpoint of the FL reversal. Also included in Figure 5 is an explicit comparison to the expected scaling of thermally activated noise caused by the secondary effect of Joule heating by the bias current. The dependence of the noise on magnetic configuration as shown in Figure 2 is already inconsistent with thermally activated spin waves, or mag-noise [22-24]. The current dependence in Figure 5 further excludes the possible interpretation in terms of mag-noise. In the absence of Joule heating, thermally activated resistance noise is independent of current because the fluctuations are driven by a constant thermal power, and the bias current simply acts as a passive probe of the thermal fluctuations [30]. The dc resistance implies that the electron temperature varies from 77 to 175 $^{\circ}$C between 10 and 24 mA, respectively. However, even when this increase in temperature is taken into account, the expected scaling of thermally activated noise falls well below the measured current dependence for both current directions.

Figures 4 and 5 also illustrate that the noise amplitude is generally larger for positive currents than for negative currents, which is the likely reason why the saturation in noise amplitude happens for only the positive current direction. This general trend is typical for the spin valves measured to date. Figure 6 shows another representation of the data that plots the resistance noise versus magnetic configuration. This figure compares the noise amplitude measured for the same magnetic configuration of the FL with respect to the RL, as inferred by the dc resistance. The dynamics stand in marked contrast to the dc resistance in which there is no difference in the magnetoresistance for the two different current directions. This dependence of noise amplitude on current direction is perhaps due to a breakdown of the two assumptions that only the FL is fluctuating and that the net field acting on the FL is the same for both current directions. Quantitative micromagnetic modeling of the data is currently underway to distinguish between ordinary magnetic effects and possible asymmetry in the spin transfer torque due to asymmetry in the leads and thin film structure of these CPP spin valves [33, 34]. However, the experimental data alone indicate that this asymmetry in noise amplitude with respect to current direction is likely due to more than just a difference in the magnetic fields acting on the FL. Using the current induced peak as a diagnostic, the peak frequency tends to be higher for negative currents than for positive currents, which suggests that a larger net magnetic field is acting on the FL and the angular excursions of the FL magnetization will be suppressed. Nevertheless, even when the peak frequency for positive currents is tuned with a magnetic field so that it coincides with the frequency for negative currents, positive currents still produce more noise.

In a separate set of experiments, spin transfer induced noise has been observed down to much slower time scales in time domain voltage measurements. Figure 7 shows two time traces for the CPP spin valve acquired with the FL biased into a low resistance state, approximately 6 Oe away from the midpoint of the FL reversal. Additional time traces have been acquired with different bias currents and magnetic fields, and the data exhibit the same qualitative current dependence as measured at high frequencies. The noisy waveform measured at +20 mA fluctuates about many different resistance values and the power spectral density [35] exhibits $1/f$ behavior with an amplitude that is within an order of magnitude of the $1/f$ noise measured at high frequencies [36]. This provides further support that the orthogonal magnetic state exhibits $1/f$ noise at microwave frequencies and shows that spin transfer can impact the magnetization dynamics at frequencies well below that of the spin precession. The existence of spectral power down to such low frequencies means that the RMS resistance noise will be even larger than that estimated from the high frequency noise spectra alone. Integrating the noise spectral density over this much larger frequency range provides a further check on the validity of the hypothesis about the origin of the resistance noise. The noise power spectral density, $S_V$, is assumed to be of the form $S_V = C \cdot I_b^2 / f$, where $C$ is a constant, $I_b$ is the bias current, and $f$ is the frequency. The data indicate that $C \approx 5 \times 10^{-5} \ \Omega^2$. Integrating this over the nine decades of frequency in which it is experimentally observed yields an RMS resistance noise of 32 m$\Omega$ [37]. This is an enormous fraction of the full $\Delta R$ of 62 m$\Omega$, yet it is still consistent with the hypothesis of the resistance noise originating from FL dynamics and GMR.

## IV. Conclusion

To summarize, we observe experimental evidence for magnetization dynamics induced by spin momentum transfer in CPP spin valves. These current induced dynamics are manifested in the noise spectra as a pronounced peak and as lower frequency noise that can either be rather flat and featureless or have $1/f$ character. The current induced noise is observable for current densities above $|j| \geq 7 \times 10^7$ A/cm$^2$ and for those combinations of current direction and magnetic configuration in which spin transfer acts to reorient the free layer in the direction opposite to that set by the net magnetic field. Otherwise, the device noise cannot be measured above the noise floor of the electronics.

The current induced noise is a function of the amplitude and direction of the bias current through the device. In addition, the amplitude and spectral characteristics of the noise are dependent upon the relative angle between the magnetizations of the free and fixed magnetic layers. A well-defined spectral noise peak appears for positive (negative) current when the free layer magnetization is close to either a parallel (antiparallel) configuration. This peak disappears and the noise spectrum is increasingly dominated by low-frequency $1/f$ noise when the free layer is driven towards an orthogonal configuration with respect to the fixed layer.

The data presented in this article provide further support for the spin torque model and the notion that a dc current can induce a state of persistent magnetization precession. However, we currently lack information about the coherence times of these oscillations because our measurements are in the frequency domain. Further measurements in the time domain are required to determine the coherence times of these persistent oscillations.

Lastly, we have identified a noise mechanism that must be considered when pursuing applications of CPP technology for devices such as read sensors in magnetic recording heads. This noise originates from magnetization fluctuations in a GMR device and, for a given set of magnetization fluctuations, the amplitude of the resulting resistance noise will scale in direct proportion to the amplitude of the GMR. For the very large current densities used in these experiments, spin transfer induced noise is the only measurable noise mechanism and dominates over Johnson noise and thermally activated magnetic noise. This is in contrast to existing GMR technology where the bias current flows parallel to the planes of the thin film multilayer and where the two predominant noise mechanisms are Johnson noise and thermally activated spin waves, or mag-noise.

We wish to acknowledge A. Rebei for many valuable technical discussions. We also want to acknowledge R. J. M. van de Veerdonk, P. A. A. van der Heijden, and T. M. Crawford for insightful comments, A. R. Eckert and J. S. Jayashankar for their technical contributions, and R. E. Rottmayer for the overall support of this work.

**Figure captions**

*Figure 1.*  Four probe resistance data from a CPP spin valve.  Field sweeps from negative to positive values are represented by a solid line.  Reversed field sweeps are indicated by a dotted line.  (a)  Resistance versus magnetic field applied along the easy axis.  (b)  Free layer minor loop for different bias currents and hard axis field values.  The resistance is referenced to its -600 Oe easy axis field value, which are 4.23, 4.23, and 4.40 Ω for curves 1, 2, and 3, respectively.  Lastly, curves 1, 2, and 3 respectively correspond to bias currents of 4, 4, and 18 mA and hard axis field values of 0, 1375, and 1375 Oe.

*Figure 2.*  (Color online) Power spectral density as a function of bias current direction and magnetic configuration.  (a)  Noise spectra acquired with a +20 mA bias current and taken at different points on the free layer minor loop, as described in the text.  The spectra taken at 34 and -10 Oe are offset by 5 and 10 $nV^2$/Hz, respectively.  (b) and (c) Power spectral density as a function of frequency and easy axis magnetic field, $H_{EA}$, for bias currents of -20 and +20 mA.  Also shown is the corresponding two-probe resistance indicating the relative magnetic configuration of the free layer.  All of the data were acquired with a constant 1375 Oe hard axis field applied.  The noise power is referenced to that of a 50 Ω resistor at 293 K and plotted in units of dB, with the contrast scale indicating the range of noise amplitude.

*Figure 3.*  (a)  Two-probe resistance versus easy axis field for a bias current of -20 mA and a hard axis field of 1375 Oe.  The resistance is referenced to its four-probe value of 4.62 Ω measured at $H_{EA}$=-600 Oe.  (b)  Corresponding RMS voltage noise calculated by integrating the power spectral density from 0.1 to 8 GHz.  (c)  RMS voltage noise integrated over smaller bandwidths.  The solid line represents the noise integrated from 0.1 to 4 GHz and the dashed line represents the noise integrated from 4 to 8 GHz.

*Figure 4.*  Dependence of RMS resistance noise on bias current and easy axis field.  A constant hard axis field of 1375 Oe is also applied.  The resistance noise is calculated from the voltage noise integrated over 0.1 to 8 GHz, as described in the text.  For both plots, the numbers indicate the bias current. Some current values have been omitted for clarity. (a) Positive current. (b) Negative current.

*Figure 5.* Dependence of RMS resistance noise on bias current. The filled circles represent experimental data from Figure 4 at a fixed easy axis field value from the midpoint of the free layer reversal. (a) Resistance noise for positive currents and -36 Oe from the free layer reversal. (b) Resistance noise for negative currents and +14 Oe from the reversal. The solid line in both figures indicates the expected behavior for thermally activated noise.

*Figure 6.* Comparison of the noise amplitude for two currents of the same amplitude but opposite direction. The resistance noise data are from Figure 4. The bottom axis shows the corresponding resistance that represents the magnetic configuration of the free layer. $\Delta R$=0 corresponds to parallel alignment and $\Delta R$=65 m$\Omega$ corresponds to antiparallel alignment as inferred from the magnetoresistance measured with +18 mA and no hard axis field.

*Figure 7.* Time traces acquired with the device in a fixed magnetic configuration, as described in the text. (a) Data for -20 mA. (b) Data for +20 mA. Note that the same vertical scale is used for (a) and (b).

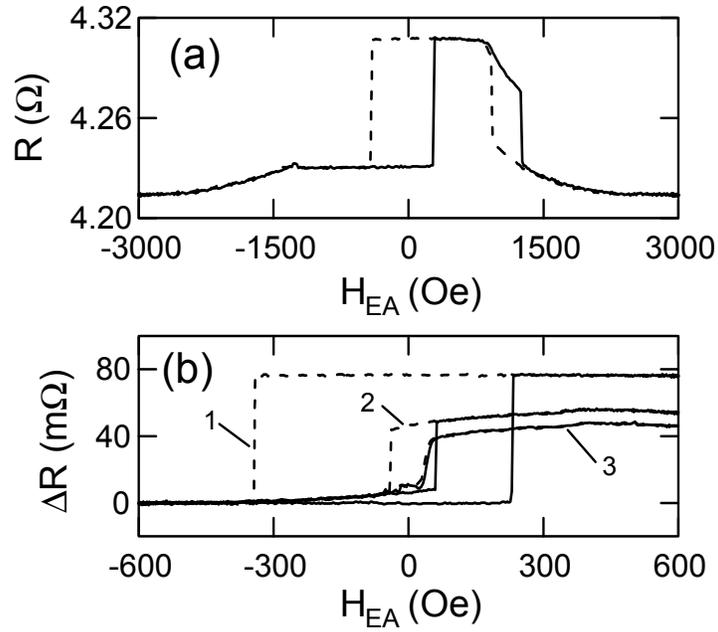

Figure 1

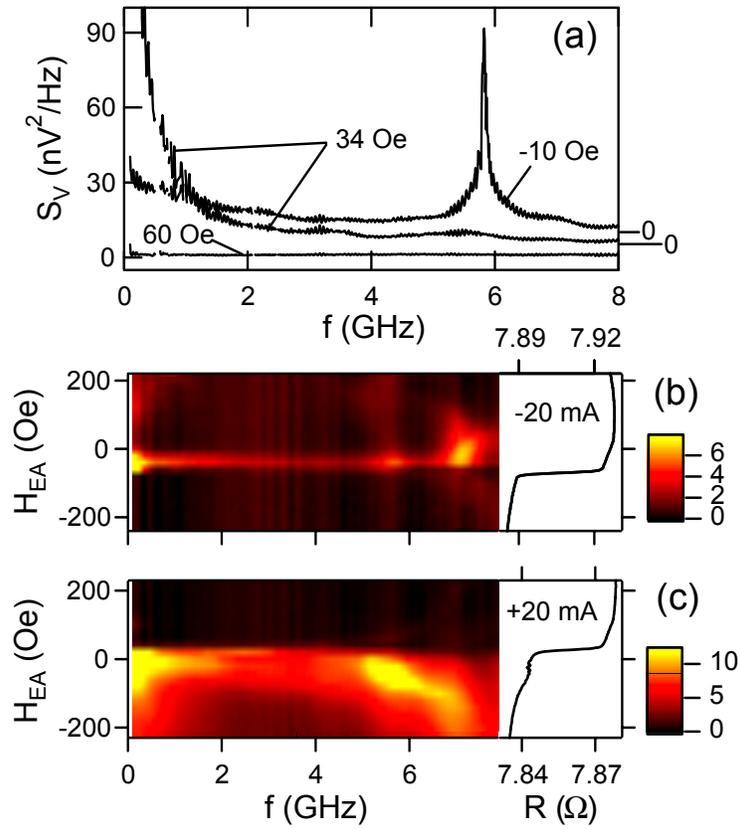

Figure 2

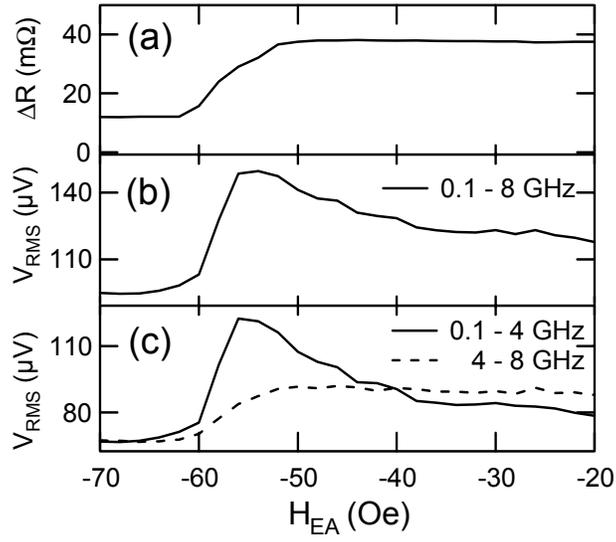

Figure 3

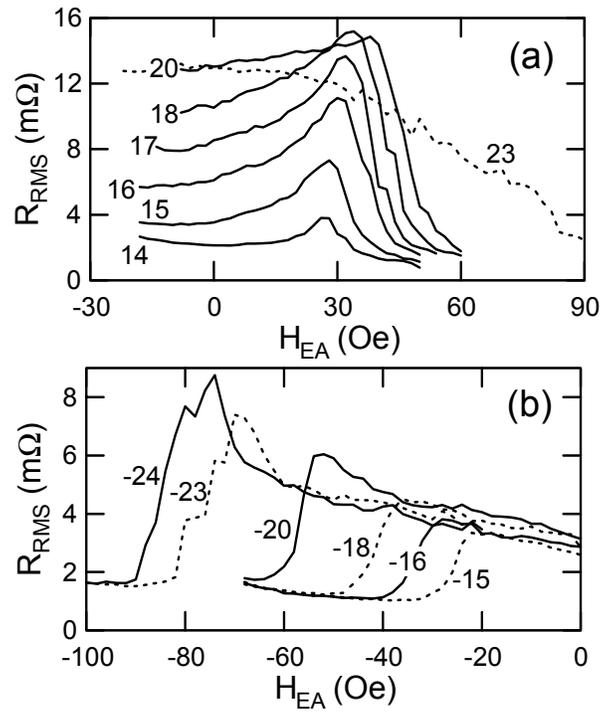

Figure 4

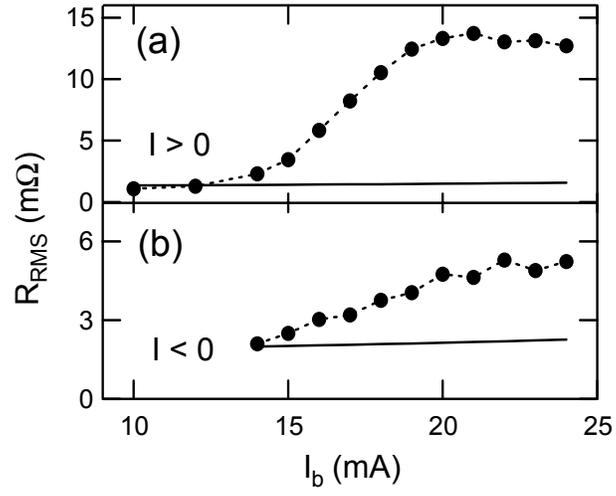

Figure 5

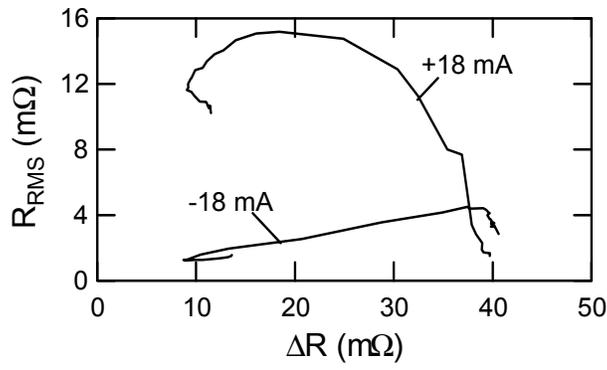

Figure 6

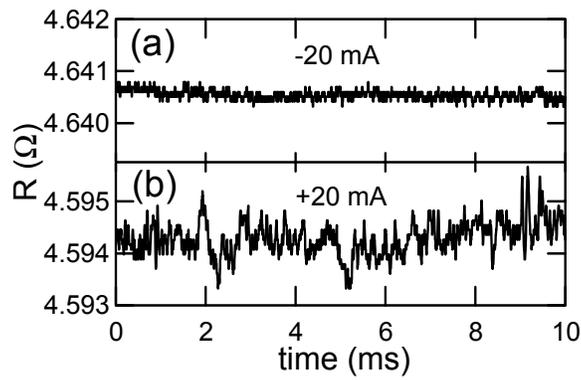

Figure 7